\documentclass[aps,pra,superscriptaddress,amsmath,amssymb,preprintnumbers,twocolumn]{revtex4}
\usepackage{amssymb}
\usepackage{graphicx}
\usepackage{bm}
\usepackage{color}
\usepackage{cancel}
\usepackage{slashed}
\usepackage{subfigure}

\newcommand{\Ignore}[1]{}

\newcommand{\Ket}[1]{\left\vert #1\right\rangle}
\newcommand{\Bra}[1]{\left\langle #1\right\vert}

\newcommand{\ii}{\mathrm{i}}
\newcommand{\ee}{\mathrm{e}}

\begin{document}

\title{Three-state Landau-Zener model in the presence of dissipation}

\author{Benedetto Militello}
\address{Dipartimento di Fisica e Chimica, Universit\`a degli Studi di Palermo, Via Archirafi 36, I-90123 Palermo, Italy}
\address{I.N.F.N. Sezione di Catania, Via Santa Sofia 64, I-95123 Catania, Italy}

\begin{abstract}
A population transfer based on adiabatic evolutions in a three-state system undergoing an avoided crossing is considered. The efficiency of the process is analyzed in connection with the relevant parameters, bringing to light an important role of the phases of the coupling constants. The role of dissipation is also taken into account, focusing on external decays that can be described by effective non-Hermitian Hamiltonians. Though the population transfer turns out to be quite sensitive to the decay processes, for very large decay rates the occurrence of a Zeno-phenomenon allows for restoring a very high efficiency. 
\end{abstract}

\maketitle

\section{Introduction}\label{sec:introduction}

Time-dependent Hamiltonians are difficult to solve, except for special classes associated with low dimensionality~\cite{ref:Barnes2012,ref:Simeonov2014,ref:Sinitsyn2018} or particular commutation properties~\cite{ref:Sriram2005,ref:Owusu2010,ref:Chrusc2015}. Beyond such special cases, only approximated resolutions are possible, based, for example, on perturbation treatments.
An important class of time-dependent Hamiltonians is that of slowly varying ones, leading to adiabatic evolutions~\cite{ref:Messiah}, which play an important role in quantum technologies, because of high efficiency in the manipulation of microscopic systems. 

In the realm of adiabatic evolutions, Landau-Zener-Majorana-Stueckelberg (LZMS)~\cite{ref:Landau,ref:Zener,ref:Majo,ref:Stuck} processes are very important. They are characterized by a two-state system with bare energies linearly changing with time, which would cross at some instant if it wasn't for an interaction between the two states which removes the degeneracy, leading to an avoided crossing. 
From the original formulation of the problem, several variants and generalizations of the model have been proposed, studied and experimentally realized. An important point that has been taken into account is the finite duration of a real experiment~\cite{ref:Vitanov1996,ref:Vitanov1999a}, in place of an infinite time interval from $-\infty$ to $\infty$ as in the original mathematical formulation. Nonlinear time-dependence of the bare energies of the two states has been considered~\cite{ref:Vitanov1999b}, as well as the avoided crossing for a two-state system whose dynamics is governed by nonlinear equations~\cite{ref:Ishkhanyan2004} or a non-Hermitian Hamiltonian~\cite{ref:Toro2017}.
Another interesting variant is the hidden crossing model, where neither the bare energies nor the dressed ones cross~\cite{ref:Fishman1990,ref:Bouwm1995}. The complementary situation is given by the total crossing model, where both the bare and the dressed energies cross~\cite{ref:Militello2015a}.

The problem of the evolution of a quantum system in the presence of a multi-level crossing has been analyzed for the first time by Majorana in his seminal work~\cite{ref:Majo}, where a spin-$j$ immersed in a magnetic field with linearly changing $z$-component has been considered. In spite of this, a complete analytical resolution of the most general multi-state LZMS model does not exist and only particular cases have been successfully analyzed. One class of these models is solvable through the so called the Independent Crossing Approximation~\cite{ref:Brundobler1993}, which is applicable because a series of independent crossings occur, each one involving two states. A very famous example is given by the Demkov-Osherov model~\cite{ref:Demkov1967}, also addressed as the equal-slope model.
A different scenario addressed as the degenerate Landau-Zener model is realized when two degenerate levels cross at the same time~\cite{ref:Vasilev2007}. 
A remarkable case characterized by a proper multi-state and multi-level crossing is the bow-tie model, where $N$ states have bare energies which cross at the same time, but the states are coupled in a particular way: one state is coupled to all the other $N-1$, while all such $N-1$ states do not couple each other. This model, originally introduced by Carroll and Hioe for the $N=3$ case~\cite{ref:Carroll1986a,ref:Carroll1986b}, has been generalized to the case of $N-2$ decoupled states crossing at the same time and two states interacting with the remaining $N-2$~\cite{ref:Demkov2000}. 
An intriguing variant of the Carroll-Hioe model has been analyzed by Ivanov and Vitanov~\cite{ref:Ivanov2008}, who considered also time-dependent coupling constants, obtaining a sort of hybrid model between LZSM and Stimulated Raman Adiabatic Passage. Other interesting studies have been developed for specific models~\cite{ref:Shytov2004,ref:Sin2015,ref:Li2017,ref:Sin2017}.
Multi-level LZMS transitions are can be related to important spin-boson models such as the time-dependent Rabi Hamiltonian~\cite{ref:Dodo2016} and Tavis-Cummings model~\cite{ref:Sun2016,ref:Sin2016}.
 
Although there are several papers dealing with dissipative adiabatic evolutions~\cite{ref:Lidar,ref:Florio,ref:ScalaOpts2011,ref:Wild2016} and, specifically, with Landau-Zener processes in the presence of interaction with the environment~\cite{ref:Ao1991,ref:Potro2007,ref:Wubs2006,ref:Saito2007,ref:Lacour2007,ref:Nel2009,ref:ScalaPRA2011}, contributions on dissipative dynamics for quantum systems with many states undergoing avoided crossings are very rare. Quite recently, Ashhab~\cite{ref:Ashhab2016} has analyzed the multi-level Landau-Zener problem in the presence of dissipation, focusing on the equal-slope and bow-tie models, and on a case with a less regular structure of the crossings that the author addresses as the triangle model.

In this paper we will consider a proper three-state avoided crossing beyond the bow-tie model, therefore including all the three couplings between the three states, also including effects of the interaction with the environment in a specific configuration of the system-bath coupling. The paper is organized as follows. 
In section \ref{sec:ideal} we study analytically the possibility of realizing a complete adiabatic population transfer, which we trace back to the presence of state whose crossings are properly avoided. 
Then we analyze the efficiency of the population transfer numerically. In section \ref{sec:dissipative} we consider the effects of the coupling with the environment in the special case where only external decays are present, meaning that the dissipation processes are responsible for incoherent transitions toward states different from the three ones we are focusing on. The relevant dynamics, which can be studied in terms of an effective non-Hermitian Hamiltonian, is extensively analyzed numerically. 
Finally, in the last section we provide some conclusive remarks.

\section{Ideal Three-State system}\label{sec:ideal}

Let us consider a three-state system described by the following Hamiltonian in the basis $\{\Ket{1}, \Ket{2}, \Ket{3}\}$: 
\begin{eqnarray} \label{eq:H3}
H_3(t) = \left(
\begin{array}{ccc}
-\kappa t & \Omega \, \ee^{\ii\phi} &  \omega  \, \ee^{\ii\varphi}  \\
 \Omega  \, \ee^{-\ii\phi}  & 0 &  \Omega  \, \ee^{\ii\phi} \\
 \omega  \, \ee^{-\ii\varphi}  &  \Omega  \, \ee^{-\ii\phi}  & \kappa t 
\end{array}
\right)\,,
\end{eqnarray} 
where $\kappa$ is changing rate of the bare energies of the first and third states, while $\Omega$ and $\omega$ are coupling constants between the bare states; here $\hbar=1$.
This kind of model is quite similar to the Carroll-Hioe one~\cite{ref:Carroll1986a,ref:Carroll1986b}, the main difference being the presence of a direct $\Ket{1}$-$\Ket{3}$ coupling.

Similarly to the two-state scenario, when $\kappa t$ is positively or negatively very large, the dressed (i.e., with interaction) eigenstates and eigenvalues approach the bare (i.e., in the absence of interaction) ones, because the coupling terms become negligible compared to the bare Bohr frequencies. In our case, the three eigenstates of the Hamiltonian $H_3(t)$ are very close to $\Ket{1}$, $\Ket{2}$ and $\Ket{3}$, with eigenvalues roughly given by $-\kappa t$, $0$ and $\kappa t$, respectively. 
In particular, for negatively very large $t$, the largest eigenvalue is approximately $-\kappa t$ and corresponds to an eigenstate roughly equal to $\Ket{1}$, while for positively large $t$ the largest eigenvalue is $\kappa t$ roughly corresponding to $\Ket{3}$.  Therefore, in the absence of crossings involving such eigenvalue and provided the Hamiltonian changes very slowly, which implies that the condition for the validity of the adiabatic following of this eigenstate is fulfilled, the state $\Ket{1}$ is mapped  into the state $\Ket{3}$, when $t$ spans an interval $[-t_0, t_0]$ with $\kappa t_0$ very large. Reciprocally, since $\Ket{3}$ is roughly equal to the negative large eigenvalue $\kappa t$ for $t$ negatively large, while for $t$ positively large the negatively large eigenvalue is $-\kappa t$ roughly equal to $\Ket{1}$, one gets that in the absence of crossings involving the lowest eigenvalue the state $\Ket{3}$ is adiabatically mapped into $\Ket{1}$. So far the situation is pretty similar to the two-state case. The only difference is that in the two-state case the crossing occurring at $t=0$ for the bare energies is removed by the coupling between the bare states. In the three-state scenario the situation is slightly more complicated. It can be proven (see Appendix \ref{app:Roots}) that for $\omega\not=\Omega$ and $2\phi-\varphi\not=m\pi$ no crossing occurs, and then the adiabatic following can be realized for sufficiently slow changes of the Hamiltonian, so that the state $\Ket{1}$ can be adiabatically mapped into $\Ket{3}$, and vice versa. For $\omega=\Omega$ and $2\phi-\varphi=2n\pi$ there is an eigenvalue --- the highest one --- which never crosses the other two. Since the corresponding eigenstate is very close to $\Ket{1}$ for negatively large values of $t$, while is essentially $\Ket{3}$ for positively large $t$, it is possible to adiabatically map the state $\Ket{1}$ into $\Ket{3}$. The opposite is not possible, because for negatively large $t$ the state $\Ket{3}$ is very close to one of the eigenstates whose corresponding eigenvalue undergoes a crossing at $t=0$. The complementary situation occurs for $\omega=\Omega$ and $2\phi-\varphi=(2n+1)\pi$: the state $\Ket{3}$ can be adiabatically mapped to $\Ket{1}$, but an efficient mapping of $\Ket{1}$ into $\Ket{3}$ is not possible. Far from this regions of the parameter space, an adiabatic population transfer can be realized in both directions: $\Ket{3}\rightarrow\Ket{1}$ and $\Ket{1}\rightarrow\Ket{3}$.

\begin{widetext}

\begin{figure}[h]
\begin{tabular}{cccl}
%\subfigure[]{\includegraphics[width=0.30\textwidth, angle=0]{f-I1.pdf}} &
%\subfigure[]{\includegraphics[width=0.30\textwidth, angle=0]{f-I2.pdf}} &
%\subfigure[]{\includegraphics[width=0.30\textwidth, angle=0]{f-I4-2.pdf}} &
\subfigure[]{\includegraphics[width=0.30\textwidth, angle=0]{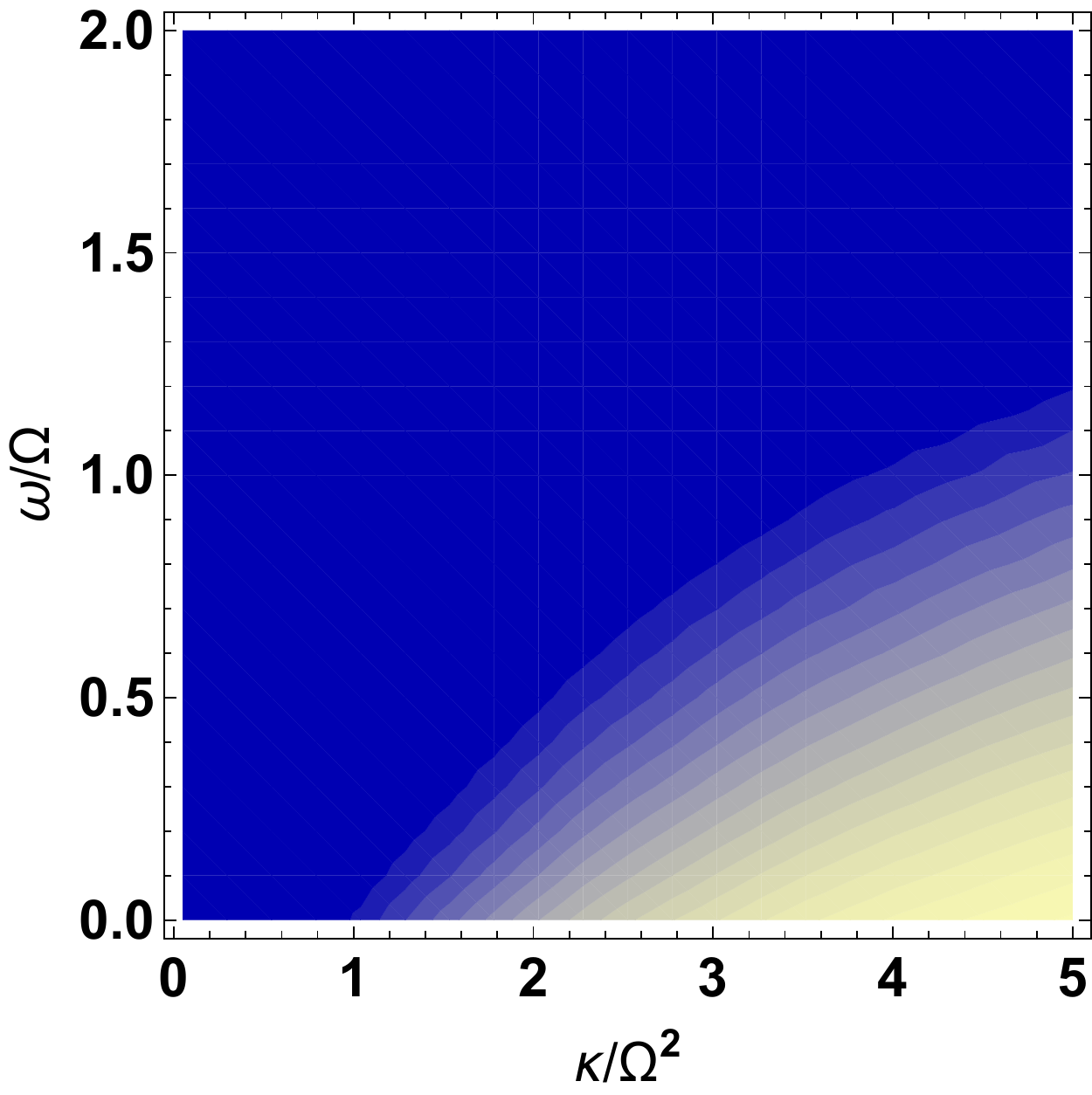}} &
\subfigure[]{\includegraphics[width=0.30\textwidth, angle=0]{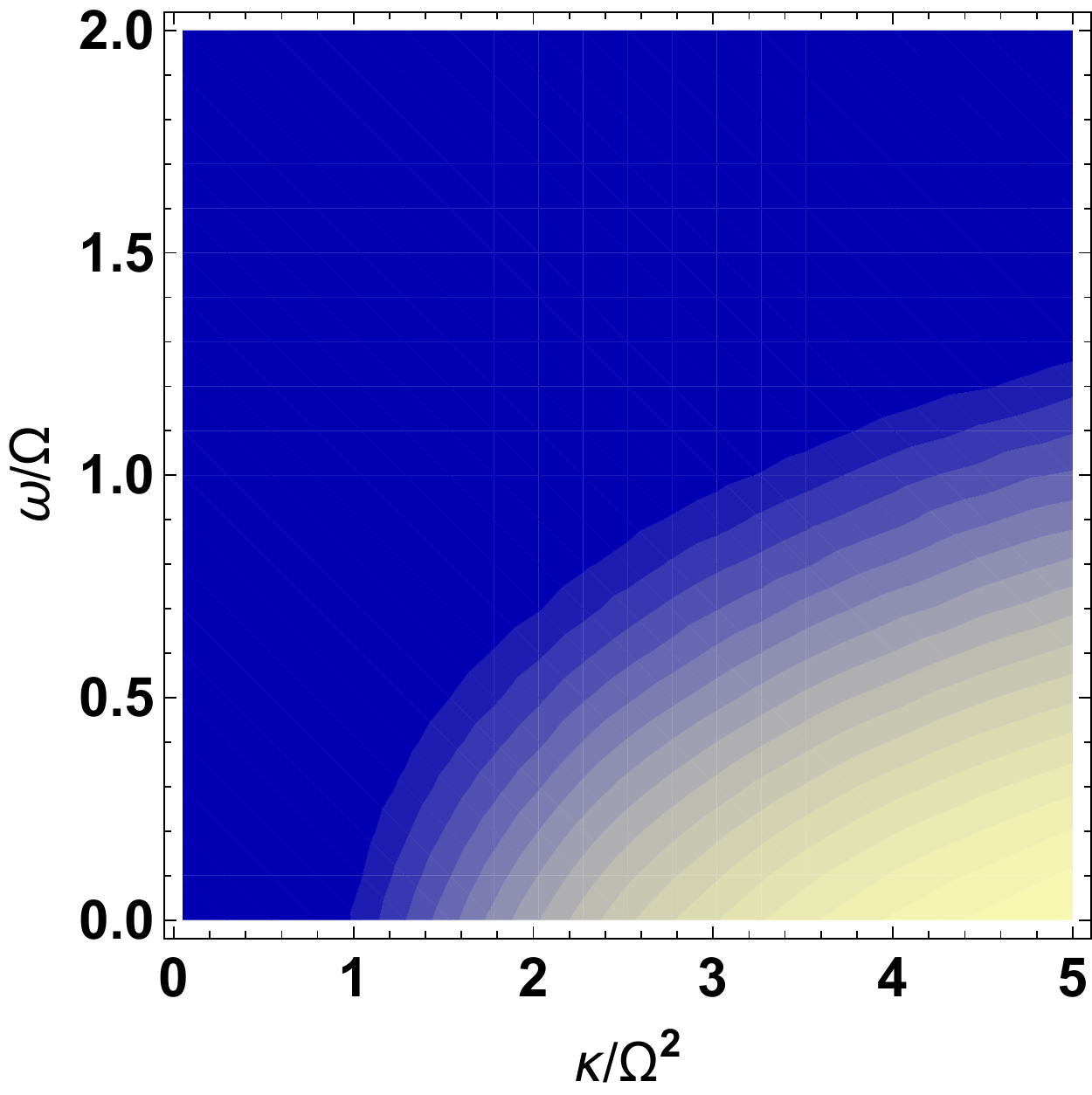}} &
\subfigure[]{\includegraphics[width=0.30\textwidth, angle=0]{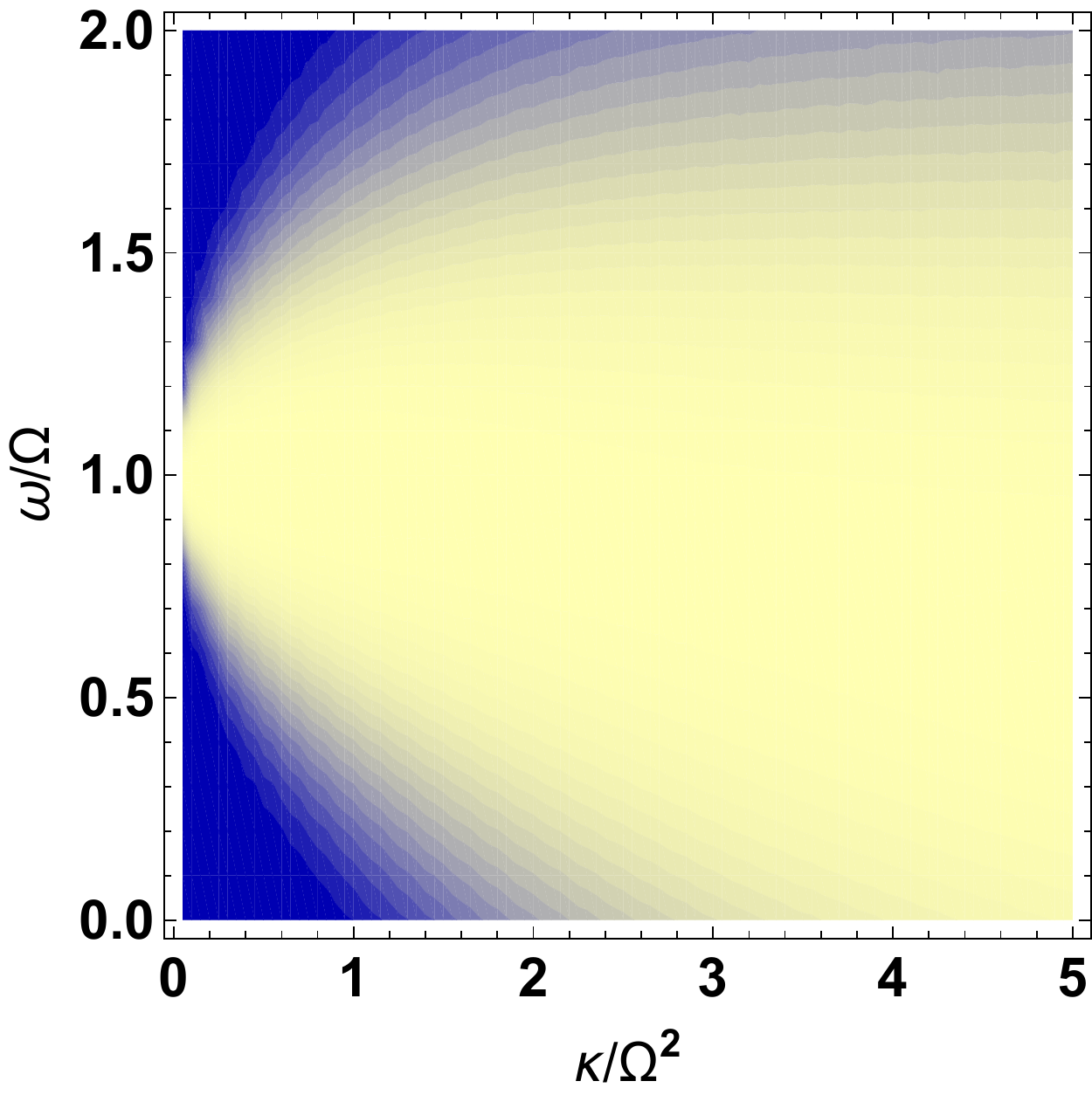}} &
\subfigure{\includegraphics[width=0.06\textwidth, angle=0]{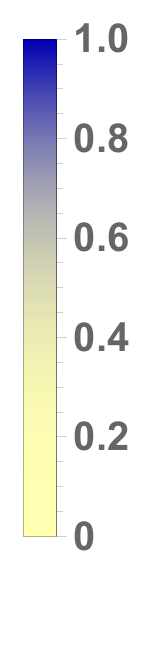}} 
\end{tabular}
\caption{Final population of the state $\Ket{3}$ when the system starts in the state $\Ket{1}$ as a function of $\omega/\Omega$ (in the range $[0,2]$) and $\kappa/\Omega^2$ (in the range $[0.05,5]$), for three different values of the phase $\varphi$: in (a) $\varphi=0$, in (b) $\varphi=\pi/2$, in (c) $\varphi=\pi$. In all three cases $\phi=0$ and $\Omega t_0=500$.} \label{fig:ideal_01}
\end{figure}

\end{widetext}

In Fig.~\ref{fig:ideal_01} are shown some examples of efficiency of the population transfer protocol based on the three-state Landau-Zener process. In particular, it is shown the final population of the state $\Ket{3}$ when the system starts in the state $\Ket{1}$ as a function of $\omega/\Omega$ and $\kappa/\Omega^2$, for three different values of the phase $\varphi$. From Fig.~\ref{fig:ideal_01}a, corresponding to $\phi=\varphi=0$, we observe that for small values of $\kappa$ the efficiency is always high, while for larger values of $\kappa$ higher values of $\omega$ are required to make the population transfer efficient. The situation for $\phi=0$ and $\varphi=\pi/2$ represented in Fig.~\ref{fig:ideal_01}b is quite similar to the $\phi=\varphi=0$ case. From  Fig.~\ref{fig:ideal_01}c it is well visible that for $\phi=0$ and $\varphi=\pi$ (implying $2\phi-\varphi=-\pi$) the efficiency becomes very low when $\omega\approx\Omega$, as expected. The region of high efficiency is very much smaller than in the other two cases.

\section{Dissipative Dynamics}\label{sec:dissipative}

Let us now introduce the effects due to the interaction with the environment. We will take into account the implications of external decays, i.e., incoherent transitions toward states different from the three on which we focused in the ideal model. To this purpose we introduce a fourth state which is considered as the ground state, so that its energy is lower than those of the other three for all $|t|<t_0$. The relevant $4\times 4$ Hamiltonian is:
\begin{eqnarray} 
H_4(t) = \left(
\begin{array}{cccc}
- \kappa t & \Omega  \, \ee^{\ii\phi} &  \omega \, \ee^{\ii\varphi} & 0 \\
 \Omega \, \ee^{-\ii\phi}  & 0 &  \Omega \, \ee^{\ii\phi} & 0 \\
 \omega \, \ee^{-\ii\varphi}  &  \Omega \, \ee^{-\ii\phi}  & \kappa t & 0 \\
 0 & 0 & 0 & -\omega_g 
\end{array}
\right)\,.
\end{eqnarray} 

Now we also assume the presence of an interaction with the environment responsible for transitions between the subspace generated by $\Ket{1},\Ket{2}, {\Ket{3}}$ and the state $\Ket{4}$:
\begin{eqnarray}\label{eq:HI}
  H_\mathrm{I} = \left( \sum_{n=1}^3 c_n  \Ket{n} \Bra{4} + h.c. \right) \sum_k g_k (a_k + a_k^\dag) \,.
\end{eqnarray}

In the case of external decays (toward lower levels  which do not belong to the subspace we are focusing on), at zero temperature, the dissipative dynamics of the main part of the system we are focusing on can be described by an effective non-Hermitian Hamiltonian (see for example Refs.\cite{ref:MilitelloPRA2010,ref:MilitelloOSID2016}). In our case, the $3\times 3$ effective non-Hermitian Hamiltonian has the following form (see appendix \ref{app:EffHam} for details):  
\begin{eqnarray} \label{eq:NonHermHeff}
\tilde{H}_3(t) = \left(
\begin{array}{ccc}
-\kappa t -\ii \Gamma_1 & \Omega  \, \ee^{\ii\phi} &  \omega  \, \ee^{\ii\varphi} \\
 \Omega  \, \ee^{-\ii\phi} & \Delta - \ii \Gamma_2 &  \Omega  \, \ee^{\ii\phi}  \\
 \omega  \, \ee^{-\ii\varphi}  &  \Omega  \, \ee^{-\ii\phi}  & \kappa t  -\ii \Gamma_3
\end{array}
\right)\,,
\end{eqnarray} 
with $\Gamma_n\propto |c_n|^2$.

In Fig.~\ref{fig:GammaCmp} is represented the final population of state $\Ket{3}$ when the system starts in $\Ket{1}$ as a function of a decay rate, for different values of the relevant parameters. In particular, in Fig.~\ref{fig:GammaCmp}a $\phi=\varphi=0$, $\omega=0$, $\kappa/\Omega^2=0.1$ and $\Omega t_0=500$ are considered. The black solid line corresponds to $\Gamma_1=\Gamma$ (the abscissa parameter) and $\Gamma_2=\Gamma_3=0$, while the red long-dashed line is related to $\Gamma_3=\Gamma$ and $\Gamma_2=\Gamma_1=0$.  The two curves perfectly match. This is well understood if one thinks that in both cases the waste of probability accumulated is roughly the same. Indeed, in the ideal case, the instantaneous eigenstate of $H_3(t)$ corresponding to the highest energy is very close to $\Ket{1}$ almost for all negative values of $t$, while it is very close to $\Ket{3}$ for all positive values of $t$. When $\Gamma_1\not=0$ and $\Gamma_2=\Gamma_3=0$, the Hamiltonian eigenstate undergoes a loss of probability roughly equal to  $\exp(-\Gamma_1 t_0)$, while for $\Gamma_3\not=0$ and $\Gamma_2=\Gamma_1=0$ the waste of probability is roughly equal to $\exp(-\Gamma_3 t_0)$.
The curve corresponding to $\Gamma_2=\Gamma$ and $\Gamma_1=\Gamma_3=0$ is different from the other two. Indeed, in this case the waste of population is accumulated mainly in a time interval around $t=0$ where the state $\Ket{2}$ is effectively involved in the dynamics, being a non negligible component of the adiabatic Hamiltonian eigenstate carrying the population. This makes the population transfer less sensitive to the decay, when $\Gamma_1=\Gamma_3=0$.

In Fig.~\ref{fig:GammaCmp}b the parameters are $\phi=\varphi=0$, $\omega=\Omega$, $\kappa/\Omega^2=0.1$ and $\Omega t_0=500$. For small values of the decay rates, the behaviors are similar to those of Fig.~\ref{fig:GammaCmp}a. A significant different behavior is instead obtained for very large values of $\Gamma_2$, where high values of efficiency are recovered. This phenomenon is traceable back to the occurrence of a Zeno-like phenomenon meant as a Hilbert state partitioning~\cite{ref:MilitelloFort2001,ref:Pascazio2002}, where a strong decay makes ineffective the couplings involving the decaying state~\cite{ref:MilitelloPRA2010,ref:ScalaPRA2011,ref:MilitelloPScr2011,ref:MilitelloQZE2011,ref:MilitelloQZE2012}. Specifically, a large value of $\Gamma_2$ implies that the state $\Ket{2}$ is essentially an eigenstate of the Hamiltonian for all values of $t$, which in turn implies that the state $\Ket{2}$ is essentially decoupled from $\Ket{1}$ and $\Ket{3}$. As a consequence, the doublet $\{\Ket{1},\Ket{3}\}$ behaves as a two-state Landau-Zener-Majorana-Stueckelberg system with coupling $\omega$ and no dissipation. Obviously, this revival of efficiency is not possible for $\omega=0$.

In Fig.~\ref{fig:GammaCmp}c a similar situation is represented, but corresponding to a shorter experiment time (with $t_0=50/\Omega$) compensated by a larger $\kappa$ to keep $\kappa t_0$ very large. Since the system is subjected to the decay for a shorter time (one tenth of the case in Fig.~\ref{fig:GammaCmp}b) the population transfer turns out to be more robust to the decay. 

Fig.~\ref{fig:DissGK} shows the population transfer efficiency as a function of $\Gamma_2$ and $\kappa$, for $\phi=\varphi=0$ and three different values of $\omega$.  In Fig.~\ref{fig:DissGK}a, where $\omega=0$, the efficiency rapidly diminishes as the decay rate increases and reaches very low values for $\Gamma_2/\Omega \sim 10^{-2}$. In Figures \ref{fig:DissGK}b and \ref{fig:DissGK}c, where $\omega\not=0$, the revival of efficiency is well visible for very large values of $\Gamma_2$.

The efficiency as a function of $\Gamma_2$ and $\omega$ is represented in the panel in Fig.~\ref{fig:DissGW1}, for a small value of $\kappa$ and three different values of phase difference $2\phi-\varphi$. For $2\phi-\varphi=0$ (a) and $2\phi-\varphi=\pi/2$ (b) there are very similar behaviors characterized again by a diminishing of the efficiency for relatively large $\Gamma_2$ and a revival for $\Gamma_2$ very large, provided $\omega$ is large enough. 
When $2\phi-\varphi=\pi$ there is also a very low efficiency even for very small $\Gamma_2$ in the region $\omega\approx \Omega$. It is interesting to note that for very large $\Gamma_2$ the revival occurs even if $\omega\approx \Omega$. This is well understood if one considers that the Hilbert space partitioning occurring for very large $\Gamma_2$ changes the interplay between the two couplings of strength $\Omega$ and the other one of strength $\omega$. Indeed, the couplings involving the state $\Ket{2}$ are made ineffective to the point that, at a first approximation, they can be considered as absent, hence removing the occurrence of a crossing. 
The panel in Fig.~\ref{fig:DissGW2} corresponds to a higher value of $\kappa$ and a smaller value of $t_0$. Though the efficiency is high in a wider range of $\Gamma_2$ values, the region of values of $\omega$ that allow for complete population transfer is smaller.

\begin{widetext}

\begin{figure}
\begin{tabular}{ccc}
%\subfigure[]{\includegraphics[width=0.30\textwidth, angle=0]{f-CA00.pdf}} &
%\subfigure[]{\includegraphics[width=0.30\textwidth, angle=0]{f-CA10.pdf}} &
%\subfigure[]{\includegraphics[width=0.30\textwidth, angle=0]{f-CA10k1.pdf}} 
\subfigure[]{\includegraphics[width=0.30\textwidth, angle=0]{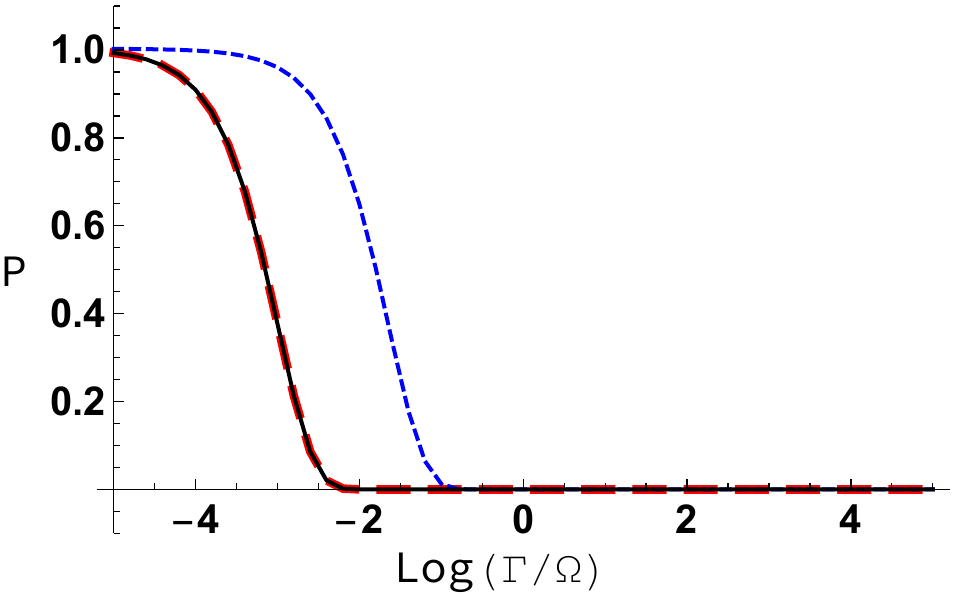}} &
\subfigure[]{\includegraphics[width=0.30\textwidth, angle=0]{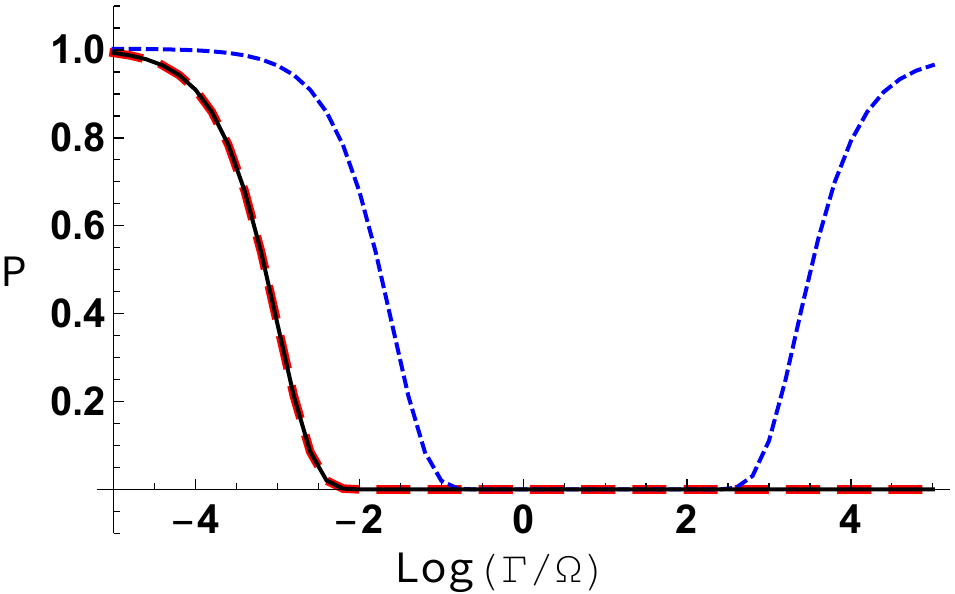}} &
\subfigure[]{\includegraphics[width=0.30\textwidth, angle=0]{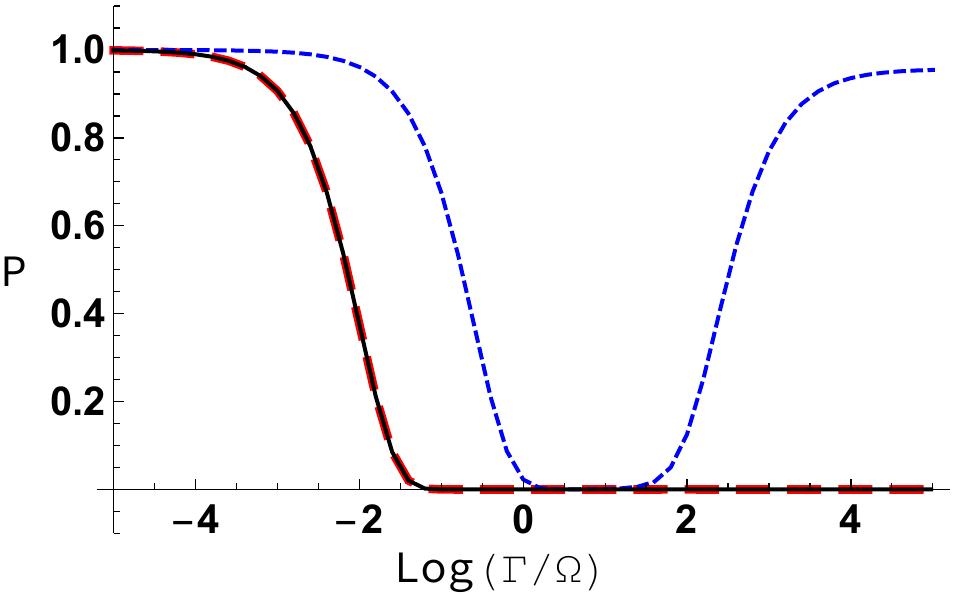}} 
\end{tabular}
\caption{Final population of the state $\Ket{3}$ when the system starts in the state $\Ket{1}$ as a function of the decay rate $\Gamma$ (in units of $\Omega$ and in logarithmic scale) for three different values of $\omega$ and $\kappa$: in (a) $\omega=0$, $\kappa/\Omega^2=0.1$ and $\Omega t_0=500$, in (b) $\omega=\Omega$, $\kappa/\Omega^2=0.1$ and $\Omega t_0=500$, in (c) $\omega=\Omega$,  $\kappa/\Omega^2=1$ and $\Omega t_0=50$.  In all the three figures, the black solid line refers to the case $\Gamma_2=\Gamma_3=0$ and $\Gamma_1=\Gamma$, the blue dashed line to $\Gamma_1=\Gamma_3=0$ and $\Gamma_2=\Gamma$, the red long-dashed line to $\Gamma_1=\Gamma_2=0$ and $\Gamma_3=\Gamma$. In all cases we have considered $\phi=\varphi=0$.} \label{fig:GammaCmp}
\end{figure}

\begin{figure}
\begin{tabular}{cccl}
%\subfigure[]{\includegraphics[width=0.30\textwidth, angle=0]{f-D00.pdf}} &
%\subfigure[]{\includegraphics[width=0.30\textwidth, angle=0]{f-D05.pdf}} &
%\subfigure[]{\includegraphics[width=0.30\textwidth, angle=0]{f-D10.pdf}} &
\subfigure[]{\includegraphics[width=0.30\textwidth, angle=0]{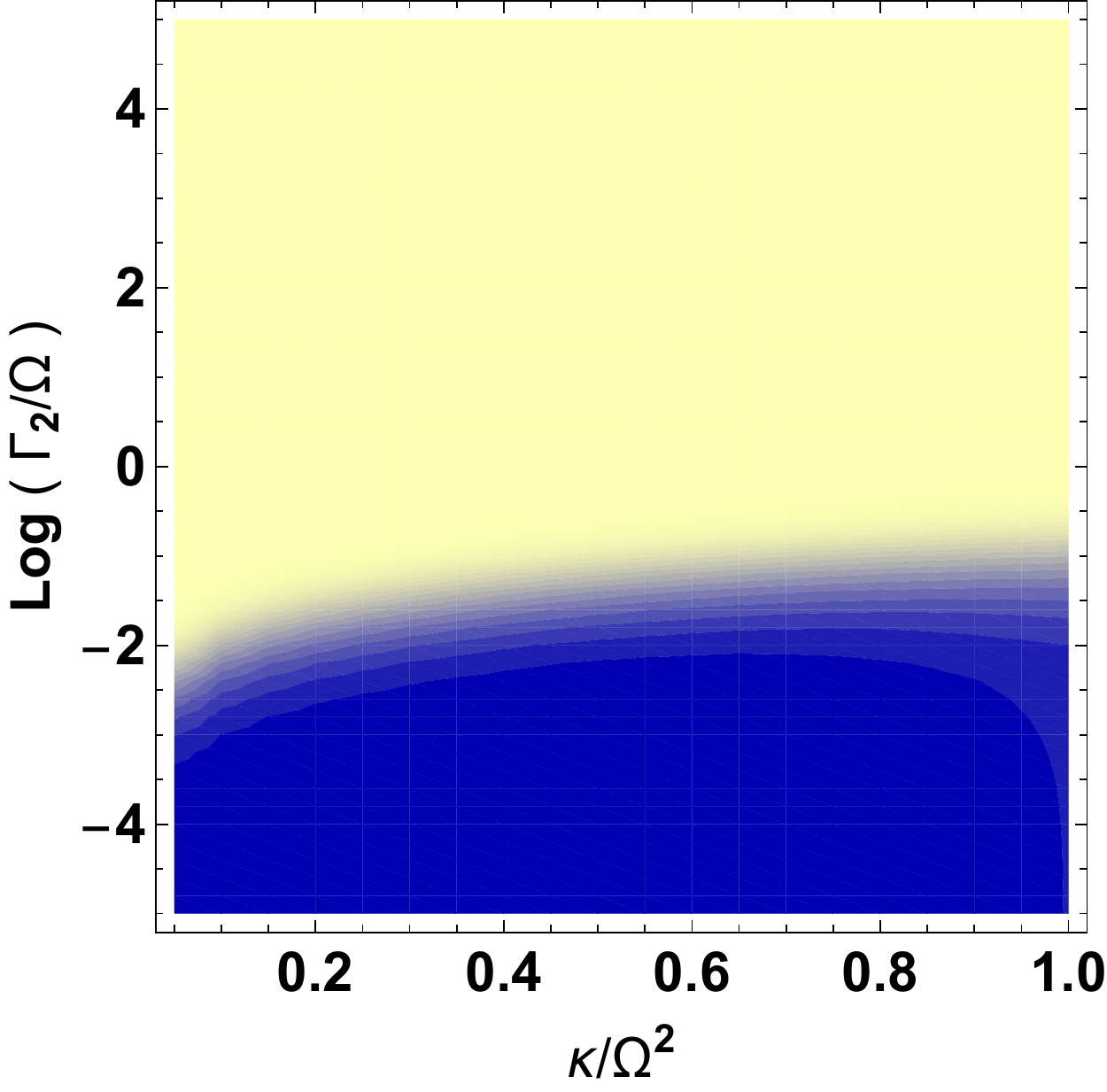}} &
\subfigure[]{\includegraphics[width=0.30\textwidth, angle=0]{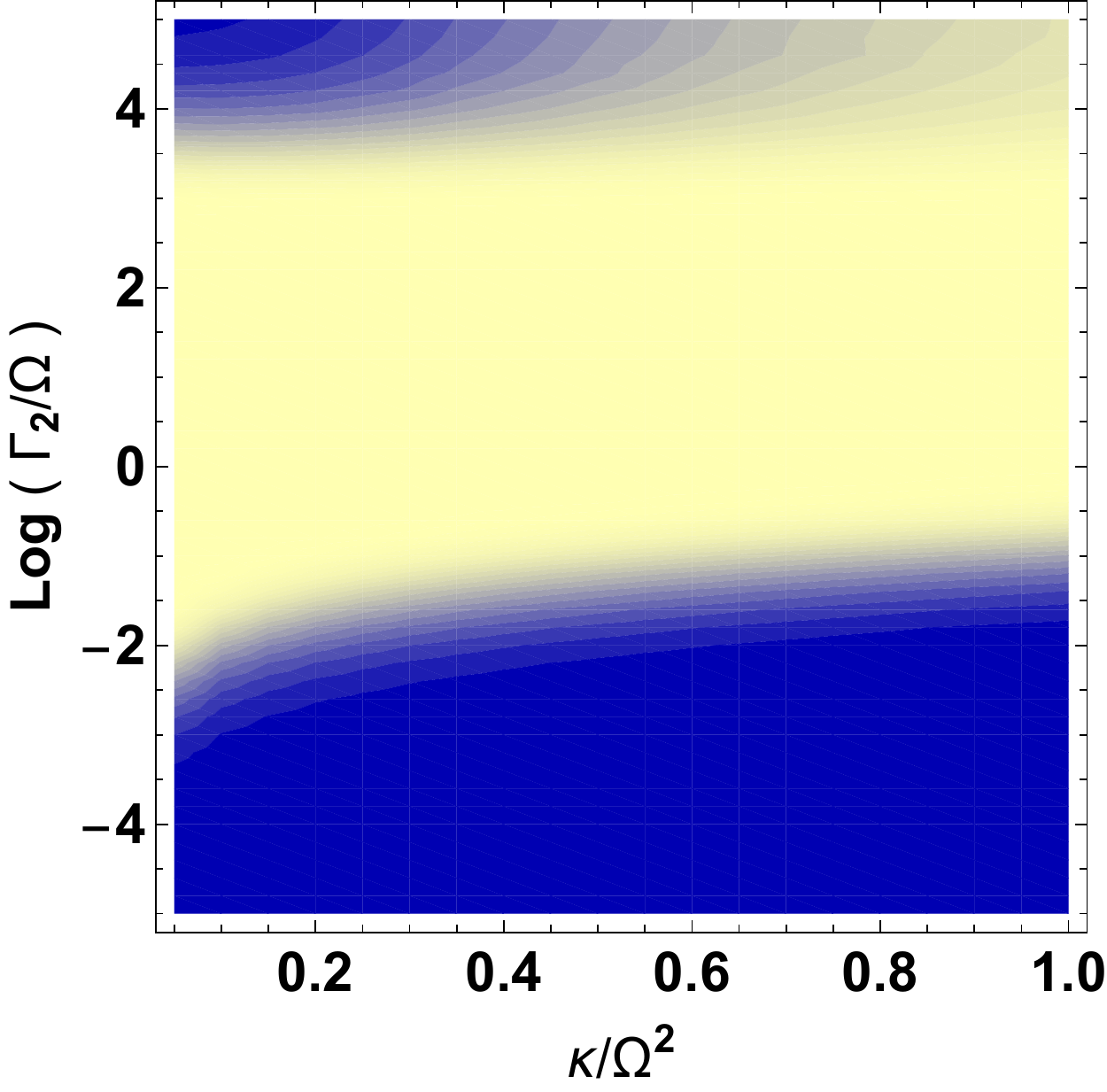}} &
\subfigure[]{\includegraphics[width=0.30\textwidth, angle=0]{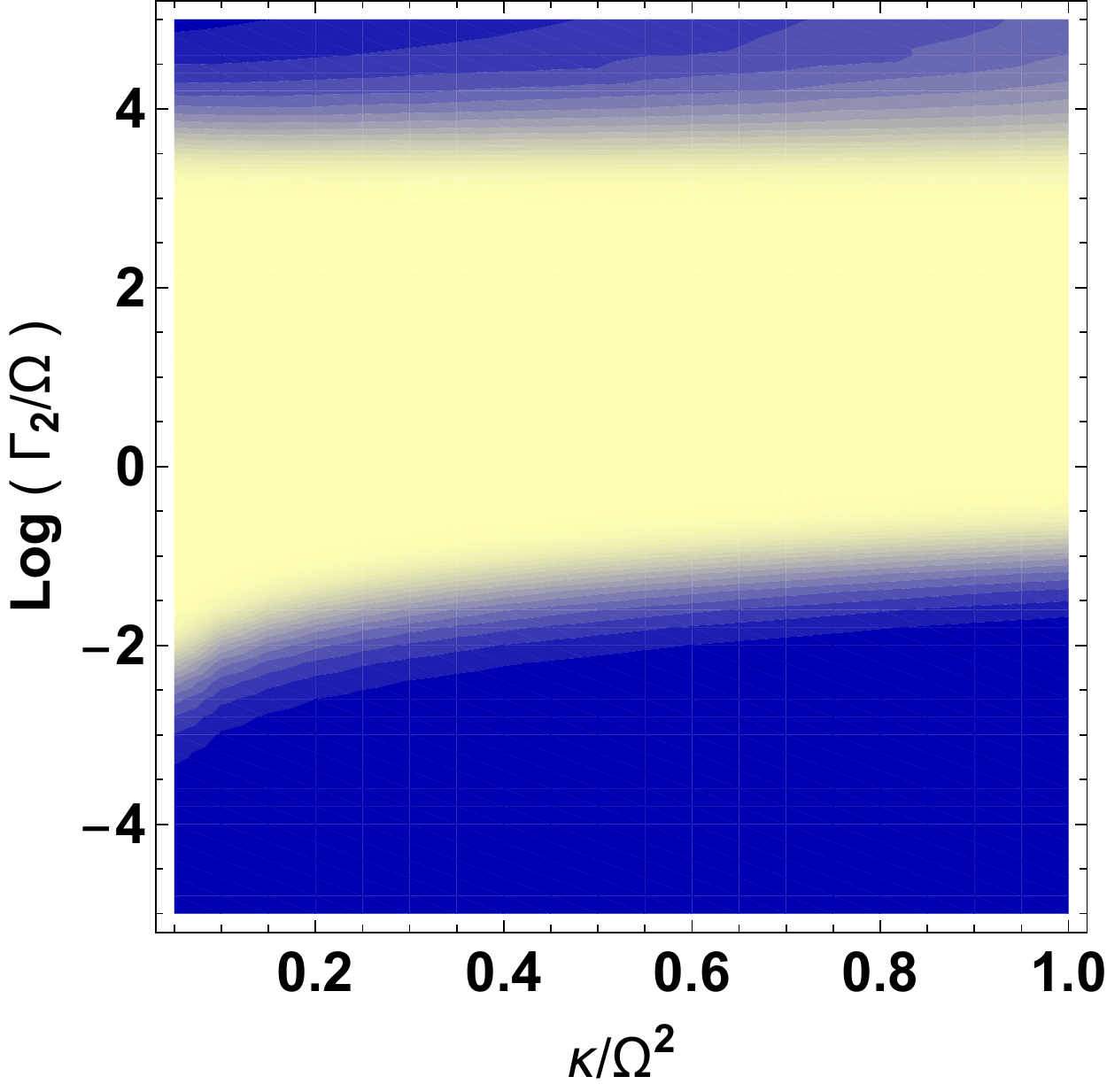}} &
\subfigure{\includegraphics[width=0.06\textwidth, angle=0]{f0.pdf}} 
\end{tabular}
\caption{Final population of the state $\Ket{3}$ when the system starts in the state $\Ket{1}$ as a function of $\mathrm{Log}(\Gamma_2/\Omega)$ (in the range $[-5,5]$) and $\kappa/\Omega^2$ (in the range $[0.05,1]$), for three different values of the phase $\omega$: in (a) $\omega=0$, in (b) $\omega=\Omega/2$, in (c) $\omega=\Omega$. In all three cases $\Gamma_1=\Gamma_3=0$, $\varphi=\phi=0$ and $\Omega t_0=500$.} \label{fig:DissGK}
\end{figure}

\begin{figure}
\begin{tabular}{cccl}
%\subfigure[]{\includegraphics[width=0.30\textwidth, angle=0]{f-E00.pdf}} &
%\subfigure[]{\includegraphics[width=0.30\textwidth, angle=0]{f-E12.pdf}} &
%\subfigure[]{\includegraphics[width=0.30\textwidth, angle=0]{f-E11.pdf}} &
\subfigure[]{\includegraphics[width=0.30\textwidth, angle=0]{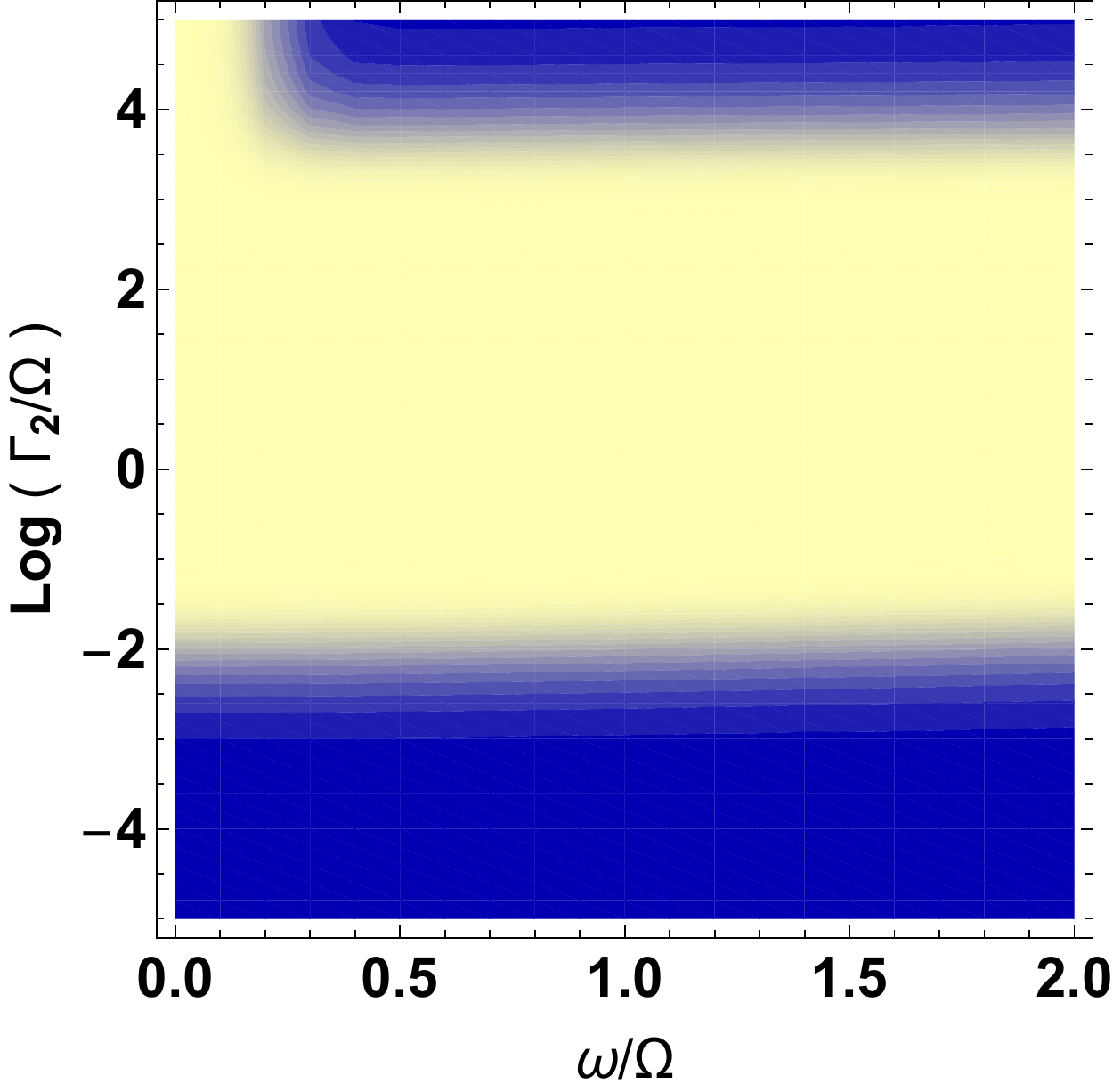}} &
\subfigure[]{\includegraphics[width=0.30\textwidth, angle=0]{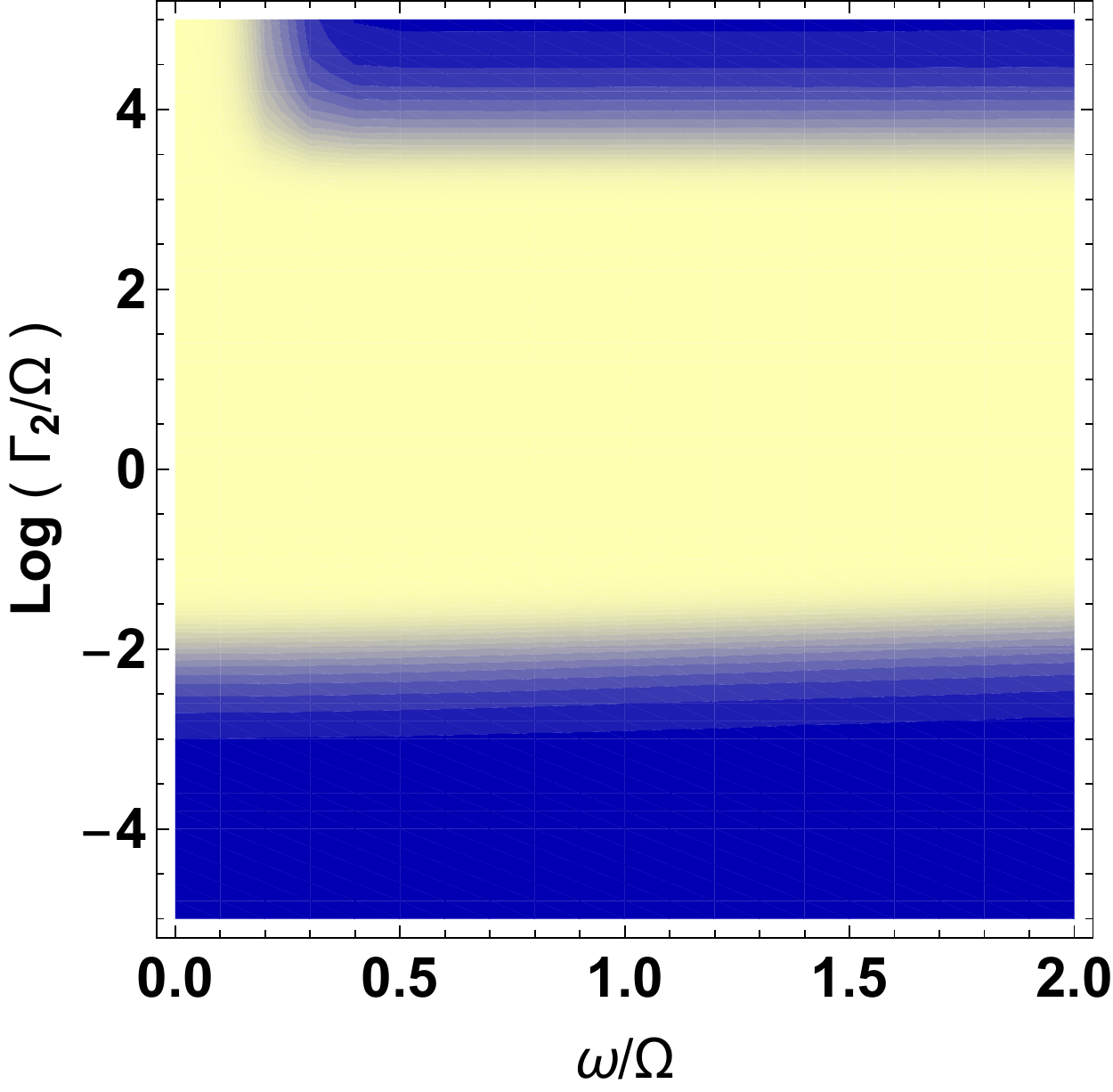}} &
\subfigure[]{\includegraphics[width=0.30\textwidth, angle=0]{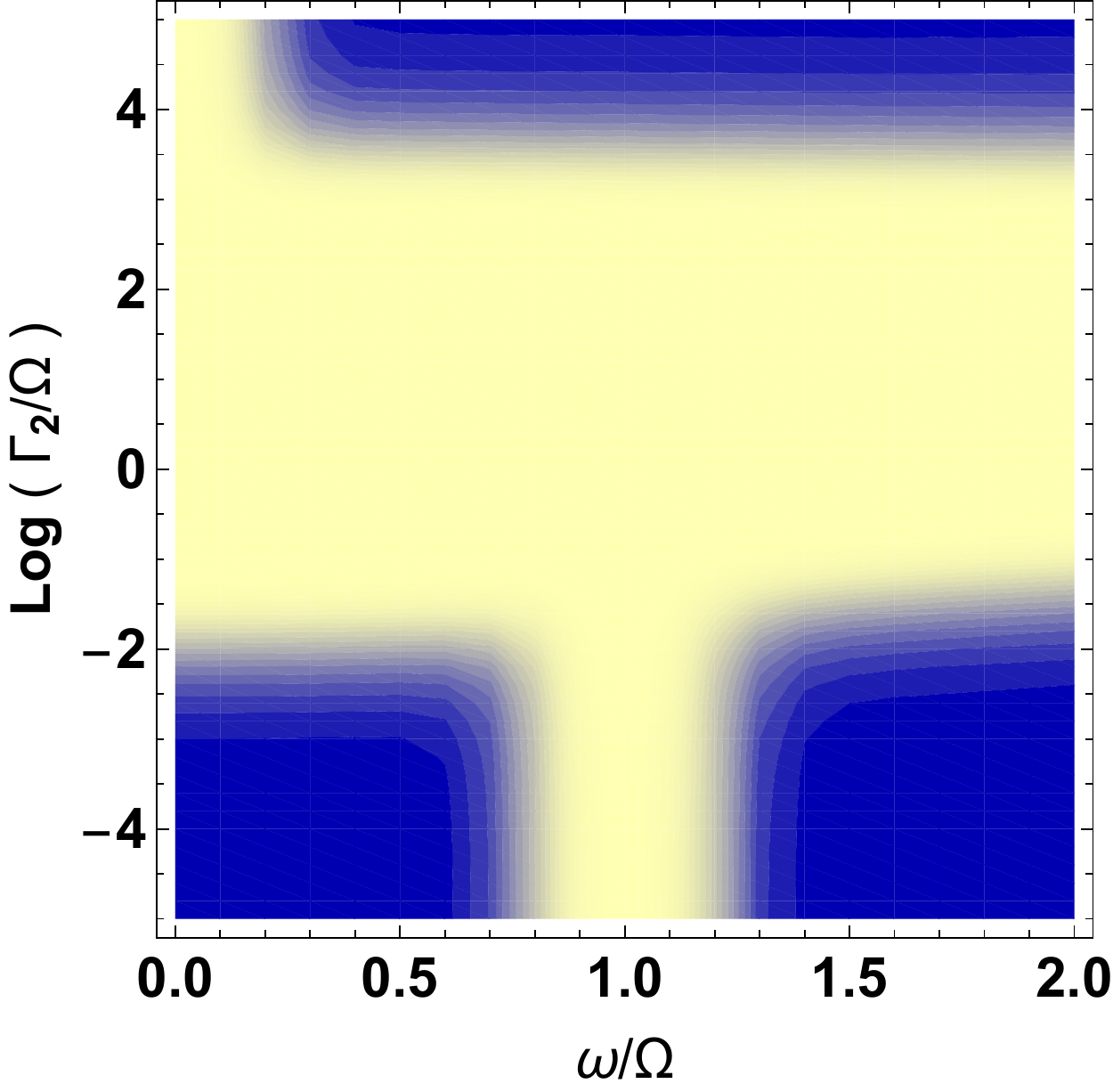}} &
\subfigure{\includegraphics[width=0.06\textwidth, angle=0]{f0.pdf}} 
\end{tabular}
\caption{Final population of the state $\Ket{3}$ when the system starts in the state $\Ket{1}$ as a function of $\mathrm{Log}(\Gamma_2/\Omega)$ (in the range $[-5,5]$) and $\omega/\Omega^2$ (in the range $[0,2]$), for three different values of the phase $\varphi$: in (a) $\varphi=0$, in (b) $\varphi=\pi/2$, in (c) $\varphi=\pi$. In all three cases $\Gamma_1=\Gamma_3=0$, $\phi=0$, $\kappa/\Omega^2=0.1$ and $\Omega t_0=500$.} \label{fig:DissGW1}
\end{figure}

\begin{figure}
\begin{tabular}{cccl}
%\subfigure[]{\includegraphics[width=0.30\textwidth, angle=0]{f-F00.pdf}} &
%\subfigure[]{\includegraphics[width=0.30\textwidth, angle=0]{f-F12.pdf}} &
%\subfigure[]{\includegraphics[width=0.30\textwidth, angle=0]{f-F11.pdf}} &
\subfigure[]{\includegraphics[width=0.30\textwidth, angle=0]{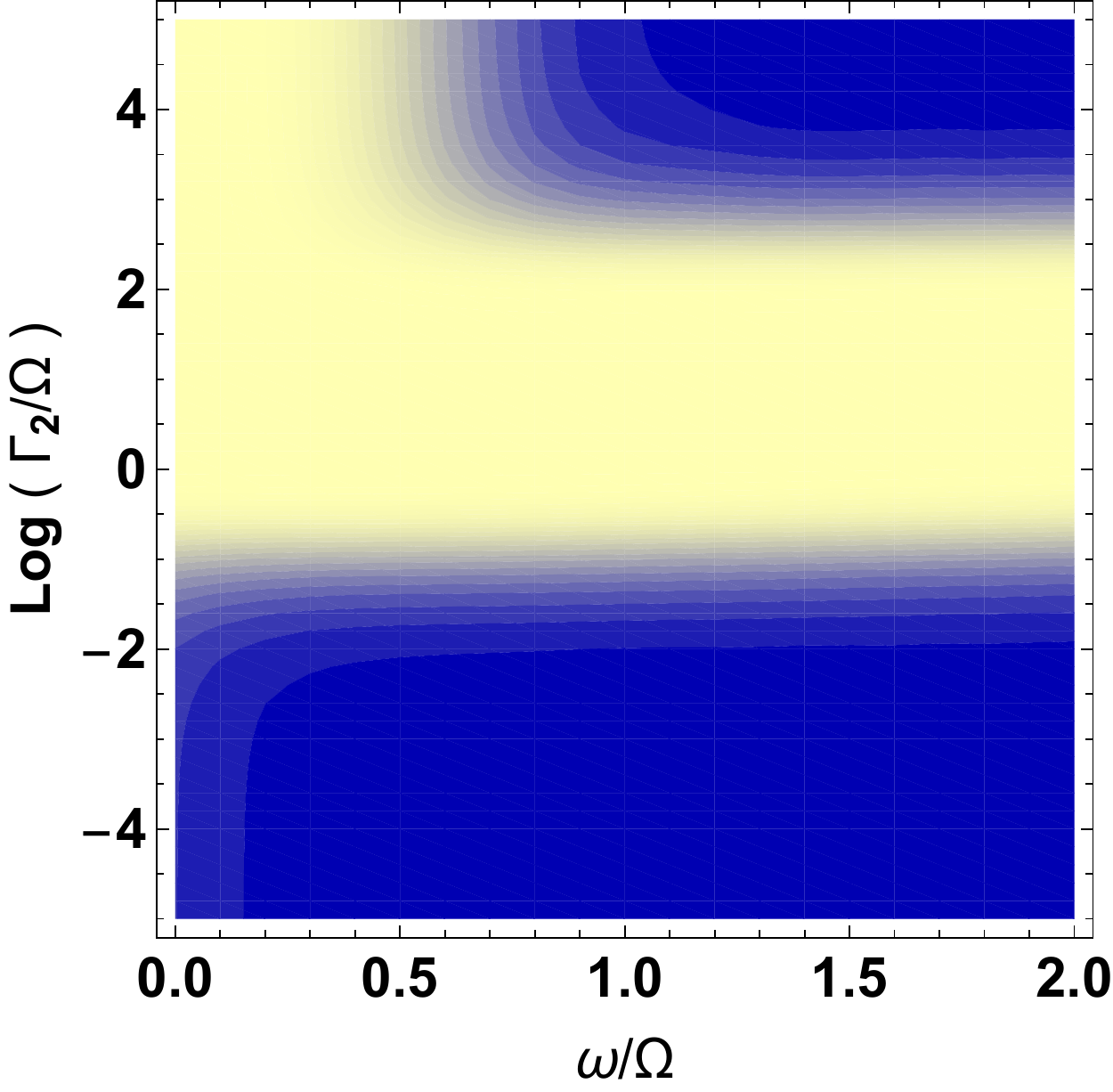}} &
\subfigure[]{\includegraphics[width=0.30\textwidth, angle=0]{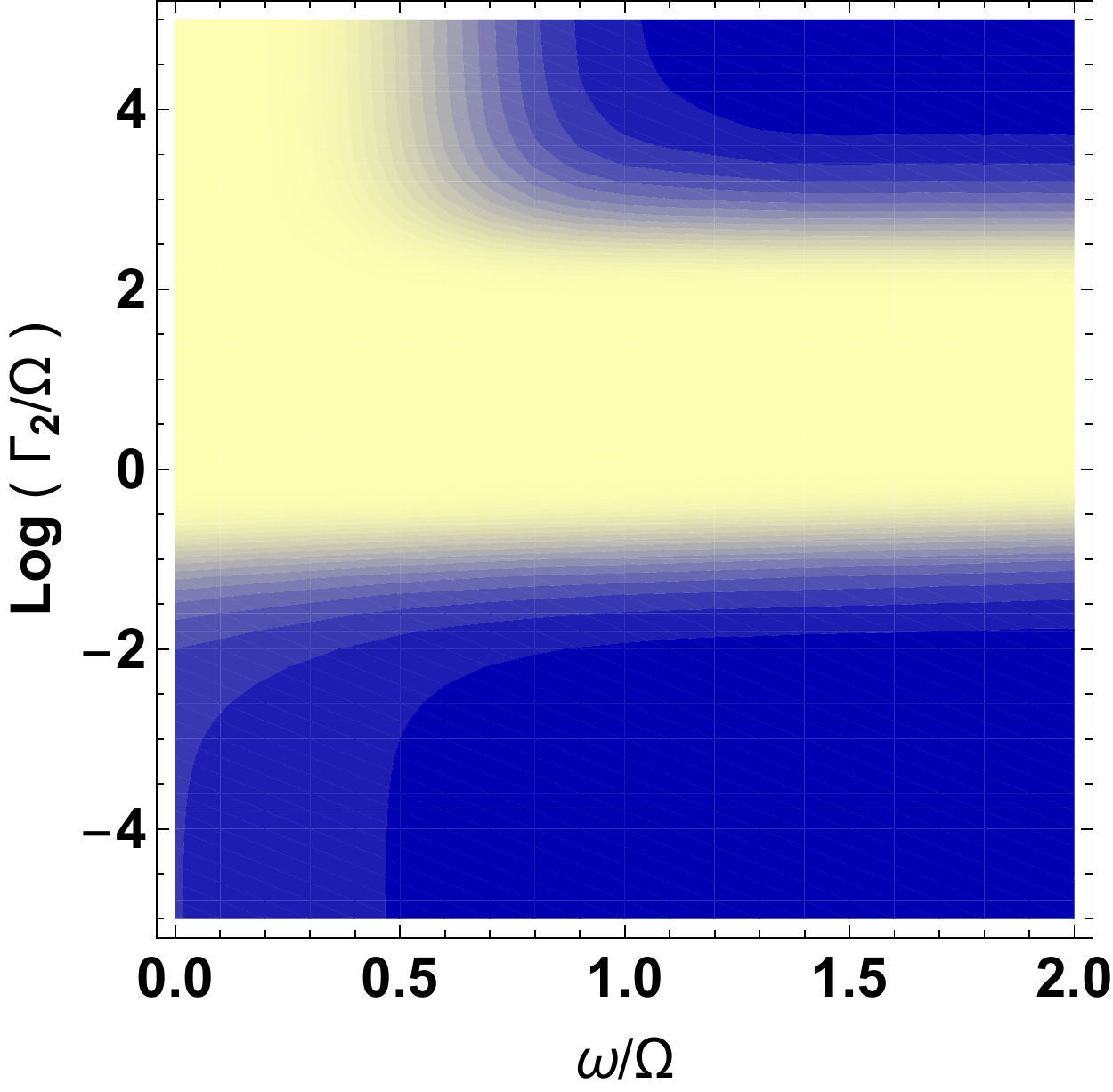}} &
\subfigure[]{\includegraphics[width=0.30\textwidth, angle=0]{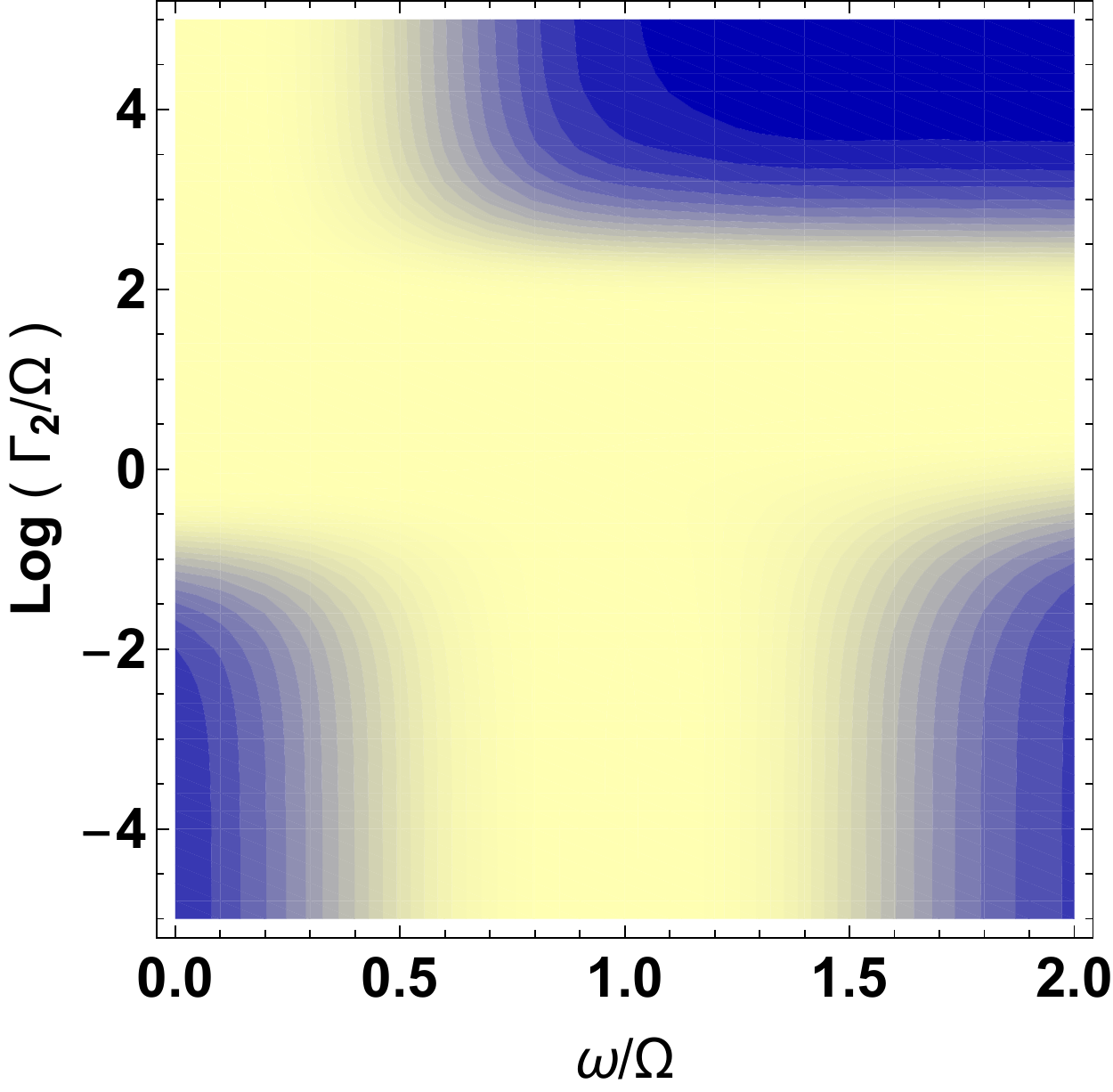}} &
\subfigure{\includegraphics[width=0.06\textwidth, angle=0]{f0.pdf}} \\
\end{tabular}
\caption{Final population of the state $\Ket{3}$ when the system starts in the state $\Ket{1}$ as a function of $\mathrm{Log}(\Gamma_2/\Omega)$ (in the range $[-5,5]$) and $\omega/\Omega^2$ (in the range $[0,2]$), for three different values of the phase $\varphi$: in (a) $\varphi=0$, in (b) $\varphi=\pi/2$, in (c) $\varphi=\pi$. In all three cases $\Gamma_1=\Gamma_3=0$, $\phi=0$, $\kappa/\Omega^2=1$ and $\Omega t_0=50$.} \label{fig:DissGW2}
\end{figure}

\end{widetext}

\section{Conclusions}\label{sec:conclusions}

In conclusion, we have analyzed a generalization of the Carroll-Hioe model which includes an additional coupling that should not be present in a bow-tie model. The constants of the two couplings between states $\Ket{1}$ and $\Ket{2}$ and between $\Ket{2}$ and $\Ket{3}$ are equal ($\Omega \ee^{-\ii\phi}$), while the third coupling between $\Ket{1}$ and $\Ket{3}$ has a different strength and phase ($\omega \ee^{-\ii\varphi}$). We have proven analytically that when $\omega\not=\Omega$ all crossings are avoided, which leads to the possibility of realizing a complete population transfer, provided the Hamiltonian changes slowly enough. When $\omega=\Omega$ the phases of the coupling constants are important, allowing for avoiding the crossing of one of the three dressed levels when $2\phi-\varphi\not=m\pi$. Otherwise, a crossing occurs which involves only two dressed states, leaving anyway the possibility of a complete adiabatic following of the state which does not cross the other two. 

The model has been analyzed also in the presence of interaction with the environment, in the special configuration where the decays happens toward states different from the three ones involved in our model, which is realized at zero temperature with system-environment interaction Hamiltonians with suitable properties. In this situation, from the microscopic model one can derive the relevant master equation, which in turn allows for identifying a non-Hermitian Hamiltonian which effectively describes the dynamics. The numerical resolution of this model has brought to light a certain sensitivity of the population transfer to the decay processes, but also the counterintuitive possibility of a recovering of the efficiency when the decay rate of state $\Ket{2}$ is very high. This behavior is traceable back to a Zeno-like phenomenon inducing a Hilbert space partitioning. In fact, the effect of the strong decay is to separate state $\Ket{2}$ from the others, hence neutralizing the $\Ket{1}$-$\Ket{2}$ and $\Ket{2}$-$\Ket{3}$ couplings. This leaves the system under the action of the sole $\Ket{1}$-$\Ket{3}$ coupling, turning the dissipative three-state LZMS model into an ideal two-state LZMS model.

\appendix

\section{}\label{app:Roots}

The characteristic equation useful to find the eigenvalues of the Hamiltonian in \eqref{eq:H3} is:
\begin{eqnarray}\label{eq:app3deg}
&& \lambda^3  - (2\Omega^2 + \omega^2 + \kappa^2 t^2)\lambda - 2\Omega^2\omega\cos\chi = 0\,,
\end{eqnarray}
with $\chi \equiv 2\phi-\varphi$.

Because of the missing $\lambda^2$-term, this equation is never of the form $(\lambda-\lambda_1)^3=0$, so that no triple root is possible. Moreover, this equation has not a double root, then never being of the form  $(\lambda-\lambda_1)^2(\lambda-\lambda_2) = 0$, unless the following three conditions occur simultaneously: $\omega=\Omega$, $\chi=m\pi$ and $t=0$. Indeed, since $(\lambda-\lambda_1)^2(\lambda-\lambda_2)=\lambda^3 - (2\lambda_1+\lambda_2)\lambda^2 + (\lambda_1^2+2\lambda_1\lambda_2)\lambda - \lambda_1^2\lambda_2$, identification of this polynomial with the left-hand side of \eqref{eq:app3deg} leads to the following conditions:  (i) $0 = \lambda_2+2\lambda_1$,   (ii) $- (2\Omega^2 + \omega^2 + \kappa^2 t^2) = \lambda_1^2+2\lambda_1\lambda_2$, and (iii) $2\Omega^2\omega\cos\chi = \lambda_1^2 \lambda_2$. After substituting $\lambda_2=-2\lambda_1$ into (ii) and (iii), and identifying the relevant two expressions for $\lambda_1$, we get the following equation: 
\begin{eqnarray}\label{eq:app_condition_1}
\kappa^2 t^2 = -\Omega^2 [x^3 - 3(\cos^2\chi)^{1/3} x + 2]\,,
\end{eqnarray}
where $x=(\omega^2/\Omega^2)^{1/3}$. 

Since $3\Omega(\cos^2\chi)^{1/3} x \le 3\Omega x$ for $x\ge 0$, and $x^3 - 3x + 2 = (x+2)(x-1)^2$, we get:
\begin{eqnarray}\label{eq:app_condition_2}
\kappa^2 t^2 \le -\Omega^2 (x+2)(x-1)^2\,.
\end{eqnarray}

Now, since the right-hand side is strictly negative for $-2<x<1$ or $x>1$, and zero for $x=-2,1$, we infer that for $\omega^2/\Omega^2 = 1$ and $\chi=m\pi$, a crossing can occur at $t=0$, while for all the other possible values of $\omega^2/\Omega^2$, from $0$ to $\infty$, no crossing is possible. Concerning the crossing occurring at $t=0$ for $\omega=\Omega$ and $\chi=2n\pi$, it corresponds to the eigenvalues $2\Omega$ (simple root) and $-\Omega$ (double root). Since no other crossing is possible, there is an eigenvalues which is always higher than the other two. Conversely, the crossing occurring at $t=0$ for $\omega=\Omega$ and $\chi=(2n+1)\pi$, it corresponds to the eigenvalues $-2\Omega$ (simple root) and $\Omega$ (double root), which implies the presence of an eigenvalues always lower than the other two.
It is worth noting that when $\cos^2\chi\not=1$, the right-hand side of \eqref{eq:app_condition_1} is strictly smaller than the right-hand side of \eqref{eq:app_condition_2}, implying the absence of a crossing even for $\omega=\pm\Omega$.

We finally observe that for $\omega=0$ or $\cos\chi=0$ the $\lambda^0$-term of the secular equation vanishes, implying a simple resolution with the three roots: $\lambda=0, \pm\sqrt{2\Omega^2+\omega^2+\kappa^2 t^2}$.

\section{}\label{app:EffHam}

Le us consider a system $\mathrm{S}$ described by the Hamiltonian $H_\mathrm{S}$ and interacting with an environment $\mathrm{S}$, whose  free Hamiltonian is $H_\mathrm{A}$ and which is prepared in a thermal state at $T$ temperature $\rho^\mathrm{A}_T$. Considering an interaction Hamiltonian between the system and the environment of the form $H_\mathrm{I}=\sum_\alpha X_\alpha \otimes V_\alpha$ (with $X_\alpha$ and $V_\alpha$ all Hermitian operators), we get the following Markovian master equation \cite{ref:Petru,ref:Gardiner}:
\begin{eqnarray}\label{eq:CompleteME}
  \nonumber
  \dot\rho &=& - \ii [H_\mathrm{S}+H_\mathrm{LS}, \rho] + \sum_{\alpha, \beta} \sum_{\omega} \gamma_{\alpha\beta}(\omega)\times \\
  &\times&  \left[ X_\beta(\omega) \rho X_\alpha^\dag (\omega) - \frac{1}{2} \left\{ X_\alpha^\dag (\omega)X_\beta (\omega), \rho\right\}\right] \,,
\end{eqnarray}
with 
\begin{eqnarray}
  && \gamma_{\alpha \beta} (\omega) = \int_{-\infty}^{+\infty} \ee^{\ii\omega s} \mathrm{tr}_\mathrm{A} \left[\tilde{V}_\alpha(s) \tilde{V}_\beta (0) \rho^\mathrm{A}_T\right] \mathrm{d}s \\
  && X_\alpha(\omega) = \sum_{\epsilon'-\epsilon=\omega} \Pi_\epsilon X_\alpha \Pi_{\epsilon'}\, \\
  && H_\mathrm{S} \Pi_\epsilon = \epsilon \, \Pi_\epsilon\,,
\end{eqnarray}
where $\omega$ spans over the Bohr frequencies of the system $\mathrm{S}$, $\tilde{V}_\alpha(t)$ is the operator $V_\alpha$ in the interaction picture and $H_\mathrm{LS}$ is the Lamb-Shift Hamiltonian commuting with $H_\mathrm{S}$.

Let us make the following assumptions: (a) the bath is at zero temperature; (b) two complementary subspaces identified by the projectors $\Pi_R$ and $\Pi_Q=\mathrm{I}-\Pi_R$ exist which are not connected by $H_\mathrm{S}$; moreover, (c) the operators $X_\alpha$ do not induce transitions inside each of such subspaces and (d) every energy level in $\Pi_R$ is higher than every level in $\Pi_Q$.
The hypothesis (a) implies that in \eqref{eq:CompleteME} the sum $\sum_\omega$ can be replaced by $\sum_{\omega>0}$, since every thermal pumping is forbidden in such a case.  The hypothesis (b) translates into $\Pi_R H_\mathrm{S} \Pi_Q=\Pi_Q H_\mathrm{S} \Pi_R=0$, which implies $[H_\mathrm{S}, \Pi_R] = [H_\mathrm{S}, \Pi_Q] = 0$. Moreover, condition (c) corresponds to $\Pi_R X_\alpha \Pi_R = \Pi_Q X_\alpha \Pi_Q = 0$, from which one easily gets $\Pi_R X_\alpha(\omega) \Pi_R = \Pi_Q X_\alpha(\omega) \Pi_Q = 0$ for every $\omega$. Finally, because of condition (d) it follows that $\Pi_R X_\alpha(\omega) \Pi_Q= 0$  and $X_\alpha(\omega) \Pi_\mathrm{R} = X_\alpha(\omega)$ for every $\omega > 0$.

After introducing the restricted state,
\begin{eqnarray}
\rho^R \equiv \Pi_R \, \rho \, \Pi_R \,,
\end{eqnarray}
and by exploiting all the previous properties, we get the following equation:
\begin{eqnarray}
  \nonumber
  \dot\rho^R &=& - \ii [\Pi_R (H_\mathrm{S}+H_\mathrm{LS}) \Pi_R, \rho^R] \\
  &-& \frac{1}{2} \sum_{\alpha, \beta} \sum_{\omega>0} \gamma_{\alpha\beta}(\omega) \left\{ X_\alpha^\dag (\omega)X_\beta (\omega), \rho^R\right\} ,
\end{eqnarray}
which can be recast in the form of pseudo-Liouville equation:
\begin{eqnarray}
  \dot\rho^R &=& - \ii (H^R \rho^R -  \rho^R(H^R)^\dag)\,, 
\end{eqnarray}
with
\begin{eqnarray}
  \nonumber
  H^R &=& \Pi_R (H_\mathrm{S}+H_\mathrm{LS}) \Pi_R \\ 
  &-& \frac{\ii}{2} \sum_{\alpha, \beta} \sum_{\omega>0} \gamma_{\alpha\beta}(\omega) X_\alpha^\dag (\omega)X_\beta (\omega) \,,
\end{eqnarray}
where $\sum_{\alpha, \beta} \sum_{\omega>0} \gamma_{\alpha\beta}(\omega) X_\alpha^\dag (\omega)X_\beta (\omega)$ is an Hermitian operator, on the basis of the property $[\gamma_{\alpha\beta}(\omega)]^*=\gamma_{\beta\alpha}(\omega)$, and has non vanishing elements in only in the subspace $\Pi_R$.

It is worth noting that assumption (d) can be relaxed, and replaced by a constraint on the initial condition, which should have zero population in all the states with energy higher than those in $\Pi_R$.

Assuming a system-bath interaction Hamiltonian of the form \eqref{eq:HI}, with state $\Ket{4}$ at lower energy than all the other three states, and a system Hamiltonian $H_\mathrm{S}=-\kappa t \Ket{1}{1} + \Delta\Ket{2}\Bra{2} + \kappa t \Ket{3}\Bra{3} - \omega_g \Ket{4}\Bra{4}$, one gets:
\begin{eqnarray}
  H_\mathrm{S}^R = \Pi_R (H_\mathrm{S}+H_\mathrm{LS}) \Pi_R - \ii \sum_{n=1}^3 \frac{\gamma_n}{2} \Ket{n}\Bra{n} \,,
\end{eqnarray}
with $\gamma_n = |c_n|^2 \gamma(\epsilon_n+\omega_g)$, $\gamma(\omega)\equiv\gamma_{11}(\omega)$ (there is only an $X_\alpha$ operator) and $\epsilon_1=-\kappa t$, $\epsilon_2=\Delta$, $\epsilon_3=+\kappa t$.

After neglecting the Lamb-shifts, introducing $\Gamma_k \equiv \gamma_k / 2$ and recovering the interaction terms $H_\mathrm{S}^R \rightarrow H_\mathrm{S}^R + \sum_{n\not=m}\Omega_{nm}\Ket{n}\Bra{m}$, we obtain the $3\times 3$ Hamiltonian in \eqref{eq:NonHermHeff}.

%\section*{Acknowledgements}

\end{document}